\documentclass[12pt]{article}
\usepackage{epsf}
\usepackage{graphicx}

\title{ {\bf Semileptonic $B_{q}\rightarrow D^{\ast}_{q}l\nu $ $(q=s, d, u)$ Decays
in  QCD Sum Rules  }}
\author{\vspace{1cm}\\
 K. Azizi \thanks {e-mail:
e146342@metu.edu.tr} ,  M. Bayar \thanks {e-mail:
mbayar@newton.physics.metu.edu.tr}\\ Physics Department, Middle East
Technical University,\\06531 Ankara, Turkey }
 \date{}
\begin{document}
\setlength{\baselineskip}{24pt} \maketitle
\setlength{\baselineskip}{7mm}
\begin{abstract}
The form factors relevant to
 $B_{q}\rightarrow D^{\ast}_{q}(J^{P}=1^{-})\ell\nu$ $(q=s, d, u)$ decays are calculated in the
 framework of the three point QCD sum rules approach. The heavy quark
 effective theory prediction of the form factors as well as $1/m_{b}$ corrections to
 those form factors are obtained. A
 comparison of the results for the ratio of form factors at zero recoil limit and
 other values of $q^{2}$ with the predictions
 of the subleading Isgur-Wise form factor application for $B\rightarrow
 D^{\ast}\ell\nu$ is presented.
 The total decay width and branching
 ratio for these decays are also evaluated using the $q^2$ dependencies of these form
 factors. The results are in good agreement with the constituent quark meson model
 and existing experimental data. The $q=
 s$ case can also be detected at LHC in the near future.
 \end{abstract}
\thispagestyle{empty}
\newpage
\setcounter{page}{1}
%%%
%%%
\section{Introduction}

Semileptonic pseudoscalar $B_{q}$ decays are crucial tools to
restrict the Standard Model (SM) parameters and search for new
physics beyond the SM. These decays provide possibility to calculate
the elements of the Cabbibo-Kobayashi-Maskawa (CKM) matrix, leptonic
decay constants as well as the origin of the CP violation.

 When LHC begins to operate, a large number of $B_{q}$ mesons will
be produced. This will provide experimental framework to check the
$B_{q}$ decay channels. An important class of $B_{s}$ and entire
$B_{u,d}$ decays occur via the $b$ quark decays. Among the $b$
decays, $b\rightarrow c$ transition plays a significant role,
because this transition is the most dominant transition among $b$
decays. Some of the $B$ decay channels could be $B_{q}\rightarrow
D^{\ast}_{q}l\nu$ $(q=s, d, u)$ via $b\rightarrow c$ transition.
These decays could give useful information about the structure of
the vector $D^{\ast}_{s}$ mesons. The observation
 of two narrow resonances with charm and strangeness, $D_{sJ}(2317)$ in the $D_{s}\pi^{0}$ invariant mass distribution
 ~\cite{1}--\cite{6}, and $D_{sJ}(2460)$ in the $D_{s}^{\ast}\pi^{0}$
and $D_{s}\gamma$ mass distribution \cite{2,3,4,6,7,8}, has raised
discussions about the nature of these states and their quark
contents \cite{9,10}.
  Analysis of the $D_{s_{0}}(2317)\rightarrow
D_{s}^{\ast}\gamma$, $D_{sJ}(2460)\rightarrow
    D_{s}^{\ast}\gamma$ and $ D_{sJ}(2460)\rightarrow D_{s_{0}}(2317)\gamma$ indicates that the quark
    content of these mesons is probably $\overline{c}s$ \cite{11}.

    Form factors are central objects in studying of the semileptonic $B_{q}\rightarrow
    D^{\ast}_{q}l\nu$ decays. For the calculation of these form factors,
    we need reliable non-perturbative approaches. Among all
    non-perturbative models, the QCD sum rules has received especial
    attention since this model is based  on the QCD Lagrangian. QCD sum rules is
     a framework which connects hadronic parameters with
    QCD parameters. In this method, hadrons are represented by their interpolating
    currents taken at large virtualities.  The correlation
    function is calculated in hadrons and quark-gluon languages. The physical quantities are
     determined by matching these two representations of correlators .
     The application of sum rules has been extended
    remarkably during the past twenty years and applied for wide
    variety of problems ( For review see for example \cite{13}).

    The aim of this paper is to analyze the semileptonic $B_{q}\rightarrow
    D^{\ast}_{q}l\nu$ decays using three point QCD sum rules method.
    Note that, this problem has been studied for $B_{q}\rightarrow
    D^{\ast}_{q}l\nu$ $(q=s, d, u)$ in constituent quark meson (CQM)
    model in \cite{zhao} and for $q=d, u  (B^{0}, B^{\pm})$ cases in
    experiment \cite{Yao}. The application of subleading Isgur-Wise form factor
    for $B\rightarrow    D^{\ast}l\nu$ at heavy quark effective theory (HQET) is calculated
    in  \cite{neubert2} (see also \cite{grozin1,ovcinkov}). Present
    work takes into account the SU(3) symmetry breaking and could be considered as an extension
    of the form factors of $D\rightarrow
K^{*} e \nu$ presented in \cite{15}.

     The paper is organized as fallow: In section II, sum rules expressions
     for form factors relevant to these decays and their HQET limit and $1/m_{b}$ corrections
      are obtained.
    The numerical analysis for form factors and their HQET limit at zero recoil and other
    values of y,
     conclusion and comparison of our
    results with the other approaches are presented in section III.

%%%
%%%
\section{Sum rules for the $B_{q}\rightarrow D^{\ast}_{q}\ell\nu$ transition form factors }
The $B_q \rightarrow  D^{\ast}_{q}$ transitions occur via the
$b\rightarrow c$ transition at the quark level. At this level, the
matrix element for this transition is given by:
\begin{equation}\label{lelement}
M_{q}=\frac{G_{F}}{\sqrt{2}} V_{cb}~\overline{\nu}
~\gamma_{\mu}(1-\gamma_{5})l~\overline{c}
~\gamma_{\mu}(1-\gamma_{5}) b.
\end{equation}
To derive the matrix elements for $B_{q}\rightarrow
    D^{\ast}_{q}l\nu$ decays, it is necessary to sandwich Eq. (\ref{lelement})
between initial and final meson states. The amplitude of the
$B_{q}\rightarrow
    D^{\ast}_{q}l\nu$ decays can be written as follows:
\begin{equation}\label{2au}
M=\frac{G_{F}}{\sqrt{2}} V_{cb}~\overline{\nu}
~\gamma_{\mu}(1-\gamma_{5})l<D^{\ast}_{q}(p',\varepsilon)\mid~\overline{c}
~\gamma_{\mu}(1-\gamma_{5}) b\mid B_{q}(p)>.
\end{equation}
 The aim is to calculate the matrix element
$<D^{\ast}_{q}(p',\varepsilon)\mid\overline{c}\gamma_{\mu}(1-\gamma_{5})
b\mid B_{q}(p)>$ appearing in Eq. (\ref{2au}). Both the vector and
the axial vector part of
  $~\overline{c}~\gamma_{\mu}(1-\gamma_{5}) b~$  contribute to the
matrix element stated above. Considering Lorentz and parity
invariances, this matrix element can be parameterized in terms of
the form factors below:
\begin{equation}\label{3au}
<D^{\ast}_{q}(p',\varepsilon)\mid\overline{c}\gamma_{\mu} b\mid
B_q(p)>=i\frac{f_{V}(q^2)}{(m_{B_{q}}+m_{D^{\ast}_{q}})}\varepsilon_{\mu\nu\alpha\beta}
\varepsilon^{\ast\nu}p^\alpha p'^\beta,
\end{equation}
\begin{eqnarray}\label{4au}
< D^{\ast}_{q}(p',\varepsilon)\mid\overline{c}\gamma_{\mu}
\gamma_{5} b\mid B_{q}(p)> &=&i\left[f_{0}(q^2)(m_{B_{q}}
+m_{D^{\ast}_{q}})\varepsilon_{\mu}^{\ast}
\right. \nonumber \\
-
\frac{f_{+}(q^2)}{(m_{B_{q}}+m_{D^{\ast}_{q}})}(\varepsilon^{\ast}p)P_{\mu}
&-& \left.
\frac{f_-(q^2)}{(m_{B_{q}}+m_{D^{\ast}_{q}})}(\varepsilon^{\ast}p)q_{\mu}\right],
\end{eqnarray}
where $f_{V}(q^2)$, $f_{0}(q^2)$, $f_{+}(q^2)$ and $f_{-}(q^2)$ are
the transition form factors and $P_{\mu}=(p+p')_{\mu}$,
$q_{\mu}=(p-p')_{\mu}$. In order to calculate these  form factors,
the QCD sum rules method is applied. Initially the following
correlator is considered:
\begin{equation}\label{6au}
\Pi _{\mu\nu}^{V;A}(p^2,p'^2,q^2)=i^2\int
d^{4}xd^4ye^{-ipx}e^{ip'y}<0\mid T[J _{\nu D^{\ast}_{q}}(y)
J_{\mu}^{V;A}(0) J_{B_{q}}(x)]\mid  0>,
\end{equation}
where $J _{\nu D^{\ast}_{q}}(y)=\overline{q}\gamma_{\nu} c$ and
$J_{B_{q}}(x)=\overline{b}\gamma_{5}q$ are the interpolating
currents of $D^{\ast}_{q}$ and $B_{q} $ mesons, respectively and
 $J_{\mu}^{V}=~\overline{c}\gamma_{\mu}b $ and $J_{\mu}^{A}=~\overline{c}\gamma_{\mu}\gamma_{5}b$
 are vector and axial vector transition currents .
Two complete sets of intermediate states with the same quantum
numbers as the currents $J_{D^{\ast}_{q}}$ and $J_{B_{q}}$ are
inserted to calculate the phenomenological part of the correlation
function given in Eq. (\ref{6au}). After standard calculations, the
following equation is obtained:
\begin{eqnarray} \label{7au}
&&\Pi _{\mu\nu}^{V,A}(p^2,p'^2,q^2)=
\nonumber \\
&& \frac{<0\mid J_{D^{\ast}_{q}}^{\nu} \mid
D^{\ast}_{q}(p',\varepsilon)><D^{\ast}_{q}(p',\varepsilon)\mid
J_{\mu}^{V,A}\mid B_{q}(p)><B_{q}(p)\mid J_{B_{q}}\mid
0>}{(p'^2-m_{D^{\ast}_{q}}^2)(p^2-m_{Bq}^2)}+\cdots
\nonumber \\
\end{eqnarray}
 where $\cdots$ represents contributions coming from higher states and continuum. The matrix
 elements in Eq. (\ref{7au}) are defined as:
\begin{equation}\label{8au}
 <0\mid J^{\nu}_{D^{\ast}_{q}} \mid
D^{\ast}_{q}(p',\varepsilon)>=f_{D^{\ast}_{q}}m_{D^{\ast}_{q}}\varepsilon^{\nu}~,~~<B_{q}(p)\mid
J_{B_{q}}\mid 0>=-i\frac{f_{B_{q}}m_{B_{q}}^2}{m_{b}+m_{q}},
\end{equation}
where $f_{D^{\ast}_{q}}$ and $f_{B_{q}}$  are the leptonic decay
constants of $D^{\ast}_{q} $ and $B_{q}$ mesons, respectively. Using
Eq. (\ref{3au}), Eq. (\ref{4au}) and Eq. (\ref{8au}) and performing
summation over the polarization of the $D^{\ast}_{q}$ meson in Eq.
(\ref{7au}) the equation below are derived:
\begin{eqnarray}\label{9amplitude}
\Pi_{\mu\nu}^{A}(p^2,p'^2,q^2)&=&\frac{f_{B_{q}}m_{B_{q}}^2}{(m_{b}+m_{q})}\frac{f_{D^{\ast}_{q}}m_{D^{\ast}_{q}}}
{(p'^2-m_{D^{\ast}_{q}}^2)(p^2-m_{B_{q}}^2)}\nonumber\\ &\times&
[-f_{0}g_{\mu\nu} (m_{B_{q}}+m_{D^{\ast}_{q}})
+\frac{f_{+}P_{\mu}p_{\nu}}{(m_{B_{q}}+m_{D^{\ast}_{q}})} \nonumber
+\frac{f_{-}q_{\mu}p_{\nu}}{(m_{B_{q}}+m_{D^{\ast}_{q}})}]\\&+&
\mbox{excited states,}\nonumber\\\Pi_{\mu\nu}^{V}(p^2,p'^2,q^2)&=&
-\varepsilon_{\alpha\beta\mu\nu}p^{\alpha}p'^{\beta}\frac{f_{B_{q}}m_{B_{q}}^2}{(m_{b}+m_{s})(m_{B_{q}}+m_{D^{\ast}_{q}})}\frac{f_{D^{\ast}_{q}}m_{D^{\ast}_{q}}}
{(p'^2-m_{D^{\ast}_{q}}^2)(p^2-m_{B_{q}}^2)}f_{V}
\nonumber \\
&+&\mbox{excited states.}
\end{eqnarray}

From the QCD (theoretical) sides, $\Pi _{\mu\nu}(p^2,p'^2,q^2)$ can
also be calculated by the help of OPE in the deep space-like region
where $p^2 \ll (m_{b}+m_{q})^2 $ and $p'^2 \ll (m_{c}+m_{q})^2$.
%To obtain the sum rules for the form factors, the two different representations of
%$\Pi _{\mu\nu}(p^2,p'^2,q^2)$ are equated using spectral representation.
The theoretical part of the correlation function is calculated by
means of OPE, and up to operators having dimension $d=6$, it is
determined by the bare-loop (Fig. 1 a) and the power corrections
(Fig. 1 b, c, d) from the operators with $d=3$,
$<\overline{\psi}\psi>$, $d=4$, $m_{s}<\overline{\psi}\psi>$, $d=5$,
$m_{0}^{2}<\overline{\psi}\psi>$ and $d=6$,
$<\overline{\psi}\psi\bar \psi \psi>$. The $d=6$ operator is ignored
in the calculations. To calculate the bare-loop contribution, the
double dispersion representation for the coefficients of
corresponding Lorentz structures appearing in the correlation
function are used:
\begin{equation}\label{10au}
\Pi_i^{per}=-\frac{1}{(2\pi)^2}\int ds'\int
ds\frac{\rho_{i}(s,s',q^2)}{(s-p^2)(s'-p'^2)}+\textrm{ subtraction
terms.}
\end{equation}
\begin{figure}
\vspace*{-1cm}
\begin{center}
\includegraphics[width=10cm]{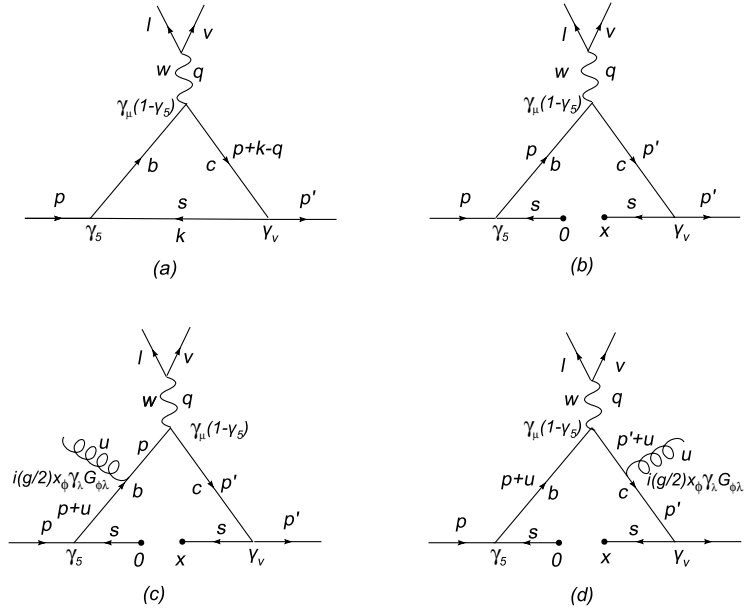}
\end{center}
\caption{Feynman diagrams for   $B_{q}\rightarrow D^{\ast}_{q}l\nu $
$(q=s, d, u)$ transitions.} \label{fig1}
\end{figure}
The spectral densities $\rho_{i}(s,s',q^2)$ can be calculated from
the usual Feynman integral with the help of Cutkosky rules, i.e., by
replacing the quark propagators with Dirac delta functions:
$\frac{1}{p^2-m^2}\rightarrow-2\pi\delta(p^2-m^2),$ which implies
that all quarks are real. After long and straightforward
calculations for the corresponding spectral densities the following
expressions are obtained:

\begin{eqnarray}\label{11au}
\rho_{V}(s,s',q^2)&=&4N_{c}I_{0}(s,s',q^2)\left[{(m_{b}-m_{q})A+(m_{c}-m_{q})B}-m_{q}\right],\nonumber\\
\rho_{0}(s,s',q^2)&=&-2N_{c}I_{0}(s,s',q^2)\Bigg[2m_{q}^{3}-2m_{q}^{2}(m_{c}+m_{b})\nonumber\\
&+&m_{q}(q^{2}+s+s'-2m_{b}m_{c})+[q^{2}(m_{b}-m_{q})\nonumber\\
&+&s(3m_{q}-2m_{c}-m_{b})+s'(m_{q}-m_{b})]A+[q^{2}(m_{c}-m_{q})\nonumber\\
&+&s(m_{q}-m_{c})+s'(3m_{q}-2m_{b}-m_{c})]B
+4(m_{b}-m_{s})C\Bigg],\nonumber \\
\rho_{+}(s,s',q^2)&=&2N_{c}I_{0}(s,s',q^2)\Bigg[m_{q}+(3m_{q}-m_{b})A+(m_{q}-m_{c})B
\nonumber
\\&+&2(m_{q}+m_{b})D+2(m_{q}-m_{b})E
\Bigg]
,\nonumber \\
\rho_{-}(s,s',q^2)&=&2N_{c}I_{0}(s,s',q^2)\Bigg[-m_{q}+(m_{q}+m_{b})A-(m_{q}+m_{c})B
\nonumber
\\&+&2(m_{q}-m_{b})D+2(m_{b}-m_{q})E
\Bigg],\nonumber \\
\end{eqnarray}
where
\begin{eqnarray}\label{12}
I_{0}(s,s',q^2)&=&\frac{1}{4\lambda^{1/2}(s,s',q^2)},\nonumber\\
 \lambda(a,b,c)&=&a^{2}+b^{2}+c^{2}-2ac-2bc-2ab,\nonumber \\
 A&=&\frac{1}{(s'+s-q^{2})^{2}-4ss'}\Bigg[(-2m_{b}^{2}+q^{^2}+s-s')s'\nonumber \\&+&m_{q}^{2}(q^{^2}-s+s')
 +m_{c}^{2}(-q^{^2}+s+s')\Bigg],\nonumber\\
 B&=&\frac{1}{(s'+s-q^{2})^{2}-4ss'}\Bigg[m_{q}^{2}(q^{^2}+s-s')\nonumber \\&+&(-2m_{c}^{2}+q^{^2}-s+s')s
 +m_{b}^{2}(-q^{^2}+s+s')\Bigg],\nonumber\\
C&=&\frac{1}{2[(s'+s-q^{2})^{2}-4ss']}\Bigg[m_{c}^{4}s+m_{b}^{4}s'\nonumber
\\&+&q^{2}[m_{q}^{4}+m_{q}^{2}(q^{^2}-s-s')+s s']
 +m_{b}^{2}m_{c}^{2}(q^{^2}-s-s')\nonumber\\
&-&(q^{^2}+s-s')s'-m_{q}^{2}(q^{^2}-s+s')\nonumber\\
&-&m_{c}^{2}m_{q}^{2}(q^{^2}+s-s')+s(q^{^2}-s+s')\Bigg],\nonumber\\
D&=&\frac{1}{[(s'+s-q^{2})^{2}-4ss']^{2}}\Bigg[m_{q}^{4}[q^{4}-2q^{2}(s-2s')+(s-s')^{2}]\nonumber
\\&+&[6m_{b}^{4}+q^{^4}+q^{^2}(4s-2s')+(s-s')^{2}-6m_{b}^{2}(q^{^2}+s-s')]s'^{2}
 \nonumber\\
&+&m_{c}^{4}[q^{^4}+s^{2}+4ss'+s'^{2}-2q^{^2}(s+s')]\nonumber\\
&-&2m_{q}^{2}s'[-2q^{^4}+(s-s')^{2}+3m_{b}^{2}(q^{^2}-s+s')+q^{^2}(s+s')]\nonumber\\
&-&2m_{c}^{2}m_{q}^{2}(q^{^2}+s^{2}+s s'-2s'^{2}+q^{^2}(-2s+s'))\nonumber\\
&+&s'[q^{^4}+q^{^2}s-2s^{2}-2q^{^2}s'+s
s'+s'^{2}+3m_{b}^{2}(-q^{^2}+s+s')]
\Bigg],\nonumber\\
E&=&\frac{1}{[s'+s-q^{2})^{2}-4ss']^{2}}\Bigg[2m_{q}^{4}q^{4}+m_{q}^{2}q^{6}-m_{q}^{4}q^{2}s-m_{q}^{2}q^{4}s-m_{q}^{4}s^{2}\nonumber
\\&-&m_{q}^{2}q^{^2}s^{2}+m_{q}^{2}s^{3}-m_{q}^{4}q^{^2}s'-m_{q}^{2}q^{^4}s'+2m_{q}^{4}s
s'
 \nonumber\\
&+&6m_{q}^{2}q^{^2}s s'+2q^{^2}s s'-m_{q}^{2}s^{2}s'-q^{^2}s^{2}s'-s^{3}s'\nonumber\\
&+&3m_{b}^{4}(q^{^2}-s+s')s'-m_{q}^{4}s'^{2}-m_{q}^{2}q^{^2}s'^{2}-m_{q}^{2}s s'^{2}\nonumber\\
&-&q^{^2}s s'^{2}+2s^{2}s'^{2}+m_{q}^{2}s'^{3}-s s'^{3}-3m_{c}^{4}s(-q^{^2}+s+s')\nonumber\\
&-&2m_{c}^{2}m_{q}^{2}[q^{^4}-2s^{2}+q^{^2}(s-2s')+s s'+s'^{2})]\nonumber\\
&+&s[q^{^2}+s^{2}+s s'-2s'^{2}+q^{^2}(-2s+s')]\nonumber\\
&+&2m_{b}^{2}\{-m_{q}^{2}(q^{^4}-2q^{^2}s+s^{2}+q^{^2}s'+s
s'-2s'^{2})\nonumber\\
&-&s'(q^{^4}+q^{^2}s-2s^{2}-q^{^2}s'+s s'+s'^{2})\nonumber\\
&+&m_{c}^{2}[q^{^4}+s^{2}+4s s'+s'^{2}-2q^{^2}(s+s')]\}
\Bigg].\nonumber\\
 \end{eqnarray}
 The subscripts V, 0 and $\pm$ correspond to the coefficients of the
 structures proportional to $i\varepsilon_{\mu\nu\alpha\beta}p'^{\alpha}p^{\beta}$, $g_{\mu\nu}$ and $\frac{1}{2}(p_{\mu}p_{\nu}
 \pm p'_{\mu}p_{\nu})$, respectively. In Eq. (\ref{11au}) $N_{c}=3$ is the number of colors.

 The integration region for the perturbative contribution
 in Eq. (\ref{10au}) is determined from the condition that arguments of the
 three $\delta$ functions must vanish simultaneously. The physical
 region in s and $s'$ plane is described by the following
 inequalities:\\
 \begin{equation}\label{13au}
 -1\leq\frac{2ss'+(s+s'-q^2)(m_{b}^2-s-m_{q}^2)+(m_{q}^2-m_{c}^2)2s}{\lambda^{1/2}(m_{b}^2,s,m_{q}^2)\lambda^{1/2}(s,s',q^2)}\leq+1.
\end{equation}

 From this inequalities, we calculate  s in terms of $s'$ in order to put to the lower
limit of integration over s. For the contribution of power
corrections, i.e., the contributions of operators with dimensions
$d=3$, $4$ and $5$, the following results were derived:
\begin{eqnarray}\label{14au}
f_{V}^{(3)}+f_{V}^{(4)}+f_{V}^{(5)}&=&\frac{1}{2}<\overline{q}q>\Bigg[-\frac{1}{rr'^{3}}
m_{c}^2(m_{0}^{2}-2m_{q}^2) \nonumber
\\&-&\frac{1}{3r^{2}r'^{2}}[-3m_{q}^2(m_{b}^2+m_{c}^2-q^{2})\nonumber
\\&+&m_{0}^{2}
(m_{b}^2+m_{b}m_{c}+m_{c}^2-q^{2})]\nonumber
\\&-&
\frac{1}{rr'^{2}}m_{c}m_{q}-
\frac{1}{r^{3}r'}m_{b}^{2}(m_{0}^{2}-2m_{q}^{2}) \nonumber
\\&+&\frac{1}{3r^{2}r'}(2m_{0}^{2}-3m_{b}m_{q}) +\frac{2}{rr'}\Bigg] ,
\nonumber \\
f_{0}^{(3)}+f_{0}^{(4)}+f_{0}^{(5)}&=&\frac{1}{4}<\overline{q}q>\Bigg[-\frac{1}{rr'^{3}}m_{c}^2(m_{0}^{2}-2m_{q}^2)
\nonumber \\
&\times&(m_{b}^{2}+2m_{b}m_{c}+m_{c}^{2}-q^{2})
\nonumber \\
&-& \frac{1}{3r^{2}r'^{2}}(m_{b}^{2}+2m_{b}m_{c}+m_{c}^{2}-q^{2})\nonumber \\
&\times& [-3m_{q}^2(m_{b}^2+m_{c}^2-q^{2})
+m_{0}^{2}(m_{b}^2+m_{b}m_{c}+m_{c}^2-q^{2})]\nonumber\\&-&
\frac{1}{3rr'^{2}}[m_{0}^{2}(m_{b}^2+3m_{b}m_{c}-q^{2})\nonumber\\&+&3(m_{c}-m_{q})m_{q}
(m_{b}^{2}+2m_{b}m_{c}+m_{c}^{2}-q^{2})]
\nonumber\\&-&\frac{1}{r^{3}r'}m_{b}^2(m_{0}^{2}-2m_{q}^{2})
(m_{b}^{2}+2m_{b}m_{c}+m_{c}^{2}-q^{2})\nonumber\\&+&\frac{1}{3r^{2}r'}
[-3(m_{b}-m_{q})m_{q}(m_{b}^{2}+2m_{b}m_{c}+m_{c}^{2}-q^{2})\nonumber\\&+&
m_{0}^{2}(m_{c}^2+3m_{b}m_{c}-q^{2})]\nonumber\\&+&\frac{1}{3rr'}(4m_{0}^{2}+
6m_{b}^{2}+12m_{b}m_{c}+6m_{c}^{2}\nonumber
\\&-&3m_{b}m_{q}+3m_{c}m_{q}-6m_{q}^{2}-6q^{2})\Bigg],
 \nonumber \\
f_{+}^{(3)}+f_{+}^{(4)}+f_{+}^{(5)}&=&\frac{1}{4}<\overline{q}q>\Bigg[-\frac{1}{rr'^{3}}
m_{c}^2(m_{0}^{2}-2m_{q}^2)\nonumber \\
&+&\frac{1}{3r^{2}r'^{2}}
[-3m_{q}^2(m_{b}^2+m_{c}^2-q^{2})\nonumber \\
&+&m_{0}^{2}(m_{b}^2+m_{b}m_{c}+m_{c}^2-q^{2})]
\nonumber \\
&+&\frac{1}{rr'^{2}}m_{c}m_{q}
<\overline{q}q>+\frac{1}{4r^{3}r'}m_{b}^2(m_{0}^{2}-2m_{q}^2)
\nonumber \\
&+&\frac{1}{3r^{2}r'}[-4m_{0}^{2}+3m_{q}(m_{b}+2m_{q})]
-\frac{1}{3rr'}\Bigg],\nonumber \\
f_{-}^{(3)}+f_{-}^{(4)}+f_{-}^{(5)}&=&\frac{1}{4}<\overline{q}q>\Bigg[-\frac{1}{rr'^{3}}
m_{c}^2(m_{0}^{2}-2m_{q}^2)\nonumber \\
&-&\frac{1}{3r^{2}r'^{2}}
[-3m_{q}^2(m_{b}^2+m_{c}^2-q^{2})\nonumber \\
&+&m_{0}^{2}(m_{b}^2+m_{b}m_{c}+m_{c}^2-q^{2})]
\nonumber \\
&-&\frac{1}{rr'^{2}}m_{c}m_{q}
-\frac{1}{r^{3}r'}m_{b}^2(m_{0}^{2}-2m_{q}^2)
\nonumber \\
&+&\frac{1}{r^{2}r'}m_{q}(-m_{b}+2m_{q})+\frac{2}{rr'}\Bigg] ,
\end{eqnarray}

where $r=p^{2}-m_{b}^{2}$ and $r'=p'^{2}-m_{c}^{2}$. Here we should
mentioned that, considering the definition of double dispersion
relation in Eq. (\ref{10au}) and parametrization of the form factors
and the coefficient of selected structures, with the changes: 1)
$b\rightarrow c$ and $c\rightarrow s$, 2) set the $m_{q}\rightarrow
0$ and 3) ignore the terms $\sim m_{s}^{2}$,  the Eqs. (\ref{11au},
\ref{14au})  reduce to the expressions for the spectral densities
and quark condensate contributions up to 5 mass dimensions for the
form factors $f_{V}$, $f_{0}$ and $f_{+}$ presented in the appendix
A of \cite{15} which describes the form factors of $D\rightarrow
K^{*} e \nu$.

 By equating the phenomenological expression given in Eq. (\ref{9amplitude}) and the
OPE expression given by Eqs. (\ref{11au}-\ref{14au}), and applying
double Borel transformations with respect to the variables $p^2$ and
$p'^2$ ($p^2\rightarrow M_{1}^2,~p'^2\rightarrow M_{2}^2$) in order
to suppress the contributions of higher states and continuum, the
QCD sum rules for the form factors $f_{V}$, $f_{0}$, $f_{+}$ and
$f_{-}$ are obtained:
\begin{eqnarray}\label{15au}
f_{i}(q^2)=\kappa\frac{(m_{b}+m_{q})
}{f_{B_{q}}m_{B_{q}}^2}\frac{\eta}{f_{D_{q}^{\ast}}m_{D_{q}^{\ast}}}e^{m_{B_{q}}^2/M_{1}^2+m_{D_{q}^{\ast}}^2/M_{2}^2}
\nonumber
\\\times[\frac{1}{(2\pi)^2}\int_{(m_{c}+m_{s})^{2}}^{s_0'} ds' \int_{f(s')}^{s_0} ds\rho_{i}(s,s',q^2)e^{-s/M_{1}^2-s'/M_{2}^2}\nonumber
\\+\hat{B}(f_{i}^{(3)}+f_{i}^{(4)}+f_{i}^{(5)})],\nonumber\\
\end{eqnarray}
 where $i=V,0$ and $\pm$, and $\hat B$ denotes the double Borel
transformation operator and $\eta=m_{B_{q}}+m_{D_{q}^{\ast}}$ for
$i=V,\pm $ and $\eta=\frac{1}{m_{B_{q}}+m_{D_{q}^{\ast}}}$ for $i=0$
are considered. Here $\kappa=+1$ for $i=\pm$ and $\kappa=-1$ for
$i=0$ and $V$. In Eq. (\ref{15au}), in order to subtract the
contributions of the higher states and the continuum, the
quark-hadron duality assumption is used, i.e., it is assumed that
\begin{eqnarray}
\rho^{higher states}(s,s') = \rho^{OPE}(s,s') \theta(s-s_0)
\theta(s'-s'_0).
\end{eqnarray}
In calculations the following rule for the double Borel
transformations is used:\\
\begin{equation}\label{16au}
\hat{B}\frac{1}{r^m}\frac{1}{r'^n}\rightarrow(-1)^{m+n}\frac{1}{\Gamma(m)}\frac{1}{\Gamma
(n)}e^{-m_{b}^{2}/M_{1}^2}e^{-m_{c}^{2}/M_{2}^2}\frac{1}{(M_{1}^{2})^{m-1}(M_{2}^{2})^{n-1}}.
\end{equation}
$~~~~~~~~~~~~~~~~~~~~~~~~~~~~~~~~~~~~~~~~~~~~~~~~~~~~~$

Here, we should mention that the contribution of higher dimensions
 are proportional to the powers of the inverse of the heavy
quark masses, so this contributions are suppressed.

Next, we present the infinite heavy quark mass limit of the form
factors for $B_{q}\rightarrow D^{\ast}_{q}l\nu $ transitions. In
HQET, the following procedure are used (see
\cite{ming,neubert1,kazem}). First, we use the following
parametrization:
 \begin{equation}\label{melau}
 y=\nu\nu'=\frac{m_{B_{q}}^2+m_{D_{q}^{\ast}}^2-q^2}{2m_{B_{q}}m_{D_{q}^{\ast}}}
 \end{equation}
 where $\nu$ and $\nu'$ are
  the four-velocities of the initial and final meson states, respectively  and $y=1$
  is so called zero recoil limit. Next, we try to find the y dependent
  expressions of the form factors by taking
  $m_{b}\rightarrow\infty$, $m_{c}=\frac{m_{b}}{\sqrt{z}}$, where
  z is given by  $\sqrt{z}=y+\sqrt{y^2-1}$ and setting the mass of light quarks to zero.
  In this limit the Borel
  parameters take the form $M_{1}^{2}=2 T_{1} m_{b}$ and $M_{2}^{2}=2 T_{2}
  m_{c}$ where $ T_{1}$ and $ T_{2}$ are the new Borel parameters.

  The new continuum thresholds $\nu_{0}$, and
$\nu_{0}'$ take the following forms in this limit
\begin{equation}\label{17au}
 \nu_{0}=\frac{s_{0}-m_{b}^2}{m_{b}},~~~~~~
 \nu'_{0}=\frac{s'_{0}-m_{c}^2}{m_{c}},
 \end{equation}
 and the new integration variables are defined as:
 \begin{equation}\label{18au}
 \nu=\frac{s-m_{b}^2}{m_{b}},~~~~~~ \nu'=\frac{s'-m_{c}^2}{m_{c}}.
 \end{equation}
 The leptonic decay constants are rescaled:
 \begin{equation}\label{21au}
\hat{f}_{B_{q}}=\sqrt{m_{b}}
f_{B_{q}},~~~~~~~\hat{f}_{D_{q}^{*}}=\sqrt{m_{c}} f_{D_{q}^{*}}.
\end{equation}
After the standard calculations, we obtain the y-dependent
expressions of the form factors as follows:

\begin{eqnarray}\label{22au}
f_{V}&=&\frac{(1+\sqrt{z})}{48
\hat{f}_{D_{q}^{*}}\hat{f}_{B_{q}}z^{1/4}}e^{(\frac{\Lambda}{T_{1}}
+\frac{\overline{\Lambda}}{T_{2}})}\Bigg\{\nonumber\\&&
\frac{3}{\pi^{2}(y+1)
\sqrt{y^{2}-1}}\int_{0}^{\nu_{0}}d\nu\int_{0}^{\nu_{0}'}d\nu'(\nu+\nu')
e^{-\frac{\nu}{2T_{1}}-\frac{\nu'}{2T_{2}}}
\theta(2y\nu\nu'-\nu^{2}-\nu'^2)\nonumber\\&+&16<\overline{q}q>\Bigg[1-\frac{m_{0}^{2}}{8}\Bigg(\frac{1}{2T_{1}^{2}}+\frac{1}{2T_{2}^{2}}
+\frac{1}{3T_{1}T_{2}}(1+\frac{1}{\sqrt{z}}+\frac{1}{z})\Bigg)\Bigg]\Bigg\},
\end{eqnarray}
\begin{eqnarray}\label{222au}
f_{0}&=&\frac{z^{1/4}}{16
\hat{f}_{D_{q}^{*}}\hat{f}_{B_{q}}(1+\sqrt{z})}
e^{(\frac{\Lambda}{T_{1}}+\frac{\overline{\Lambda}}{T_{2}})}\Bigg\{
\frac{3}{\pi^{2}
\sqrt{y^{2}-1}}\int_{0}^{\nu_{0}}d\nu\int_{0}^{\nu_{0}'}d\nu'(\nu+\nu')
e^{-\frac{\nu}{2T_{1}}-\frac{\nu'}{2T_{2}}}\nonumber\\
&&\theta(2y\nu\nu'-\nu^{2}-\nu'^2)+\frac{<\overline{q}q>\sqrt{z}}{3}\Bigg[
\Bigg(\frac{1}{2}+\frac{1}{2z}+\frac{1}{\sqrt{z}}\Bigg)\nonumber\\
&&\Bigg(16-m_{0}^{2}(\frac{1}{T_{1}^{2}}+\frac{1}{T_{1}^{2}})\Bigg)-
\frac{m_{0}^{2}}{T_{1}T_{2}}
\Bigg(1+\frac{1}{3z^{\frac{3}{2}}}+\frac{4}{3\sqrt{z}}
+\frac{1}{z}+\frac{\sqrt{z}}{3}\Bigg)\Bigg]\Bigg\},
\end{eqnarray}
\begin{eqnarray}\label{2222au}
f_{+}&=&\frac{(1+\sqrt{z})}{96
\hat{f}_{D_{q}^{*}}\hat{f}_{B_{q}}z^{1/4}}e^{(\frac{\Lambda}{T_{1}}+\frac{\overline{\Lambda}}{T_{2}})}\Bigg\{\nonumber\\
&& \frac{9}{\pi^{2}(y+1)
\sqrt{y^{2}-1}}\int_{0}^{\nu_{0}}d\nu\int_{0}^{\nu_{0}'}d\nu'(\nu+\nu')e^{-\frac{\nu}{2T_{1}}-\frac{\nu'}{2T_{2}}}
\theta(2y\nu\nu'-\nu^{2}-\nu'^2)\nonumber\\
&-&16<\overline{q}q>
\Bigg[1+\frac{m_{0}^{2}}{8}\Bigg(\frac{1}{2T_{1}^{2}}+\frac{1}{2T_{2}^{2}}
+\frac{1}{3T_{1}T_{2}}(1+\frac{1}{\sqrt{z}}+\frac{1}{z})\Bigg)\Bigg]\Bigg\},
\end{eqnarray}
\begin{eqnarray}\label{22222au}
f_{-}&=&-\frac{(1+\sqrt{z})}{96\hat{f}_{D_{q}^{*}}\hat{f}_{B_{q}}z^{1/4}}
e^{(\frac{\Lambda}{T_{1}}+\frac{\overline{\Lambda}}{T_{2}})}\Bigg\{\nonumber\\
&& \frac{9}{\pi^{2}(y+1)
\sqrt{y^{2}-1}}\int_{0}^{\nu_{0}}d\nu\int_{0}^{\nu_{0}'}d\nu'(\nu+\nu')e^{-\frac{\nu}{2T_{1}}-\frac{\nu'}{2T_{2}}}
\theta(2y\nu\nu'-\nu^{2}-\nu'^2)\nonumber\\
&+&16<\overline{q}q>\Bigg[1-\frac{m_{0}^{2}}{8}\Bigg(\frac{1}{2T_{1}^{2}}+\frac{1}{2T_{2}^{2}}
+\frac{1}{3T_{1}T_{2}}(1+\frac{1}{\sqrt{z}}+\frac{1}{z})\Bigg)\Bigg]\Bigg\},
\end{eqnarray}
where $\Lambda=m_{B_{q}}-m_{b}$ and
$\bar{\Lambda}=m_{D_{q}^{*}}-m_{c}$.

At the end of this section, we would like to present
$\frac{1}{m_{b}}$ corrections for the form factors in Eqs.
(\ref{22au})-(\ref{22222au}) using subleading Isgur-Wise form
factors similar to \cite{neubert2} (see also
\cite{neubert1,grozin}). These corrections are given as:
\begin{eqnarray}\label{3333au}
f_{V}^{(1/m_{b})}&=&\frac{m_{B}+m_{D}^{*}}{\sqrt{m_{B}m_{D}^{*}}}
\Bigg\{\frac{\Lambda}{2m_{b}}+\frac{\Lambda}{m_{b}}[\rho_{1}(y)-\rho_{4}(y)]\Bigg\},\nonumber\\
f_{0}^{(1/m_{b})}&=&\frac{(y+1)\sqrt{m_{B}m_{D}^{*}}}{m_{B}+m_{D}^{*}}
\Bigg\{\frac{\Lambda}{2m_{b}}\frac{y-1}{y+1}+\frac{\Lambda}{m_{b}}[\rho_{1}(y)-\frac{y-1}{y+1}\rho_{4}(y)]\Bigg\},\nonumber\\
f_{+}^{(1/m_{b})}&=&\frac{1}{2}f_{V}^{(1/m_{b})},\nonumber\\
f_{-}^{(1/m_{b})}&=&-f_{+}^{(1/m_{b})},
\end{eqnarray}
where the explicit expressions for $\rho_{i}(y)$ functions are given
in \cite{neubert2}. The value of those functions at zero recoil
limit $(y=1)$ are given as
\begin{eqnarray}\label{555555au}
\rho_{1}(1)=\rho_{2}(1)=0,~~~~~ \rho_{3}(1)\simeq0,~~~~~
 \rho_{4}(1)\simeq\frac{1}{3}.
\end{eqnarray}

\section{Numerical analysis}
This section is devoted by numerical analysis for the form factors
$f_{V}(q^2)$, $f_{0}(q^2)$, $f_{+}(q^2)$ and $f_{-}(q^2)$. From sum
rule expressions of these form factors it is clear that  the
condensates, leptonic decay constants of $B_{q}$ and $D_{q}^{\ast}$
mesons, continuum thresholds $s_{0}$ and  $s'_{0} $ and Borel
parameters $M_{1}^2$ and $M_{2}^2$ are the main input parameters. In
the numerical analysis the values of the condensates are chosen at a
fixed renormalization scale of about $1$ GeV. The values of the
condensates are\cite{20} :
$<\overline{u}u>=<\overline{d}d>=-(240\pm10~MeV)^3$,
$<\overline{s}s>=(0.8\pm0.2)<\overline{u}u>$ and
$m_{0}^2=0.8~GeV^2$.
 The quark masses are taken to be $ m_{c}(\mu=m_{c})=
 1.275\pm
 0.015~ GeV$, $m_{s}= 95\pm25 ~MeV$, $m_{u}=(1.5-3) ~MeV$, $m_{d}\simeq(3-5) ~MeV$ \cite{Yao} and $m_{b} =
(4.7\pm
 0.1)~GeV$ \cite{20}. The mesons masses are chosen to be $m_{D_{s}^{\ast}}=2.112~GeV$ , $m_{D_{u}^{\ast}}=2.007~GeV$ ,
  $m_{D_{d}^{\ast}}=2.010~GeV$, $m_{B_{s}} =5.3~GeV$, $m_{B_{d}} =5.2794~GeV$and $ m_{B_{u}}=5.2790~GeV$\cite{Yao}. For
 the values of the leptonic decay
constants of $B_{q}$ and $D_{q}^{\ast} $ mesons the results obtained
from two-point QCD analysis are used: $f_{B_{s}} = 0.209\pm
 38~ GeV $ \cite{13}, $f_{D_{s}^{\ast}} =0.266\pm0.032
  ~GeV $\cite{11}. For the others $f_{B_{d(u)}}=0.14\pm0.01
  ~GeV $ and $f_{D_{d(u)}^{\ast}}=0.23\pm0.02
  ~GeV $\cite{Yao} are selected. The threshold parameters
$s_{0}$ and $s_{0}' $ are also determined from the two-point QCD sum
rules: $s_{0} =(35\pm 2)~ GeV^2$ \cite{12} and $s_{0}' =(6-8)~ GeV^2
$ \cite{11}. The Borel parameters $M_{1}^2$ and $M_{2}^2 $ are not
physical quantities, hence form factors should not depend on them.
The reliable regions for the Borel parameters $M_{1}^2 $ and
$M_{2}^2$ can be determined by requiring that both the continuum
contribution  and the contribution of the operator with the highest
dimension be small. As a result of the above-mentioned requirements,
the working regions are determined to be $ 10~ GeV^2 < M_{1}^2 <25~
GeV^2 $ and $ 4~ GeV^2 <M_{2}^2 <10 ~GeV^2$.

 To determine the decay width of $B_{q} \rightarrow D_{q}^{\ast}l\nu$, the $q^2$ dependence of the form factors $ f_{V}(q^2)$, $f_{0}(q^2)$, $f_{+}(q^2)$ and $f_{-}(q^2)$ in the whole
physical region $ m_{l}^2 \leq q^2 \leq (m_{B_{q}} -
m_{D_{q}^{\ast}})^2$ are needed. The value of the form factors at
$q^2=0$ are given in Table 1.
\begin{table}[h]
\centering
\begin{tabular}{|c||c|c|c|} \hline
$f_{i}(0)$& $B_{s}\rightarrow D_{s}^{\ast}\ell \nu $ &
$B_{d}\rightarrow D_{d}^{\ast}\ell \nu$ & $B_{u}\rightarrow
D_{u}^{\ast}\ell \nu $
\\\cline{1-4}\hline\hline
$f_{V}(0)$ & $0.36\pm0.08$ & $0.47\pm0.13$ & $0.46\pm0.13
$\\\cline{1-4} $f_{0}(0)$ & $0.17\pm0.03$ & $0.24\pm0.05$ &
$0.24\pm0.05$\\\cline{1-4} $f_{+}(0)$ & $0.11\pm0.02$ &
$0.14\pm0.025$ & $0.13\pm0.025$\\\cline{1-4}$f_{-}(0)$
&$-0.13\pm0.03$& $-0.16\pm0.04$ & $-0.15\pm0.04$\\\cline{1-4}
\end{tabular}
\vspace{0.8cm} \caption{The value of the form factors at $q^2=0$}.
\label{tab:2}
\end{table}

 The $q^2 $ dependence of the form
factors can be calculated from QCD sum rules (for details, see
\cite{15,16}). To obtain the $q^2$ dependent expressions of the form
factors from QCD sum rules, $q^2 $ should be stay approximately $1~
GeV^2$ below the perturbative cut, i.e., up to $10 ~GeV^2$. Our sum
rules, also, are truncated at $\simeq10 ~GeV^{2}$, but in the
interval $0 \leq q^2 \leq 10~ GeV^2$ we can trust the sum rules. For
the reliability of the sum rules in the full physical region, the
parametrization of the form factors were identified such that in the
region $0 \leq q^2 \leq 10~ GeV^2$, these parameterizations coincide
with the sum rules prediction. Figs. \ref{fig1}, \ref{fig2},
\ref{fig3} and \ref{fig4} show the dependence of the form factors
$f_{V}(q^2)$, $f_{0}(q^2)$, $f_{+}(q^2)$ and $f_{-}(q^2)$ on $q^2$.
To find the extrapolation of the form factors, we choose the
following two fit functions.\\ i)
 \begin{equation}\label{17au}
 f_{i}(q^2)=\frac{f_{i}(0)}{1+\alpha\hat{q}+\beta\hat{q}^2+\gamma\hat{q}^3+\lambda\hat{q}^4},
\end{equation}
where $\hat{q}=q^2/m_{B_{q}}^2$. The values of the parameters
 $f_{i}(0),\alpha,\beta,\gamma$, and $\lambda$ are
given in  Tables 2, 3 and 4.\\ ii)
 \begin{equation}\label{17au}
 f_{i}(q^2)=\frac{a}{(q^{2}-m_{B^{*}}^{2})}+\frac{b}{(q^{2}-m_{fit}^{2})}.
\end{equation}
 The values for a, b and
$m_{fit}^{2}$ are given in Tables 5, 6 and 7. For details about the
fit parametrization (ii) which is theoretically  more reliable  and
some other fit functions see \cite{damir,damirinbali}. These two
parameterizations coincide well with the sum rules predictions in
the whole physical region $0 \leq q^2 \leq 10~ GeV^2$ and also for
$q^{2}<0$ region. For higher $q^{2}$, starting from the upper limit
of the physical region the two fit functions deviate from each other
and this behavior is almost the same for all form factors. As an
example, we present the deviation of above mentioned fit functions
in Fig. \ref{fig6}. From this figure, we see that in the outside of
the physical region the fit (i) growthes more rapidly than fit (ii).
The fit parametrization (ii) depicts that the $m_{B^{*}}$ pole
exists outside the allowed physical region and related to that one
could calculate the hadronic parameters such as $g_{BB^{*}D^{*}}$
(see \cite{damir,damirinbraunu}).
\begin{table}[h] \centering
\begin{tabular}{|c||c|c|c|c|c|} \hline
  & f(0)  & $\alpha$ & $\beta$& $\gamma$& $\lambda$\\\cline{1-6} \hline \hline
 $f_{V}$ & 0.38 & -2.53 & 2.77& -2.41& 0.03\\\cline{1-6}
 $f_{0}$ & 0.18  & -1.77 & 0.98& -0.23& -3.50\\\cline{1-6}
 $f_{+}$ & 0.12  & -2.90 & 3.66& -3.72& -1.69\\\cline{1-6}
 $f_{-}$ & -0.15  & -2.63 & 2.72& -0.99& -6.48\\\cline{1-6}
 \end{tabular}
 \vspace{0.8cm}
\caption{Parameters appearing in the fit function (i) for form
factors of the $B_{s}\rightarrow D_{s}^{\ast}(2112)\ell\nu$ at
$M_{1}^2=19~GeV^2$, $M_{2}^2=5~GeV^2.$} \label{tab:1}
\end{table}

\begin{table}[h]
\centering
\begin{tabular}{|c||c|c|c|c|c|} \hline
  & f(0)  & $\alpha$ & $\beta$& $\gamma$& $\lambda$\\\cline{1-6}\hline \hline
 $f_{V}$ & 0.46 & -2.90 & 2.99& 0.67& -5.04\\\cline{1-6}
 $f_{0}$ & 0.24  & -0.21 & 2.19& -1.68& -2.15\\\cline{1-6}
 $f_{+}$ & 0.13  & -4.21 & 9.52& -16.86& 12.97\\\cline{1-6}
 $f_{-}$ & -0.15  & -3.93 & -8.03& -13.48&9.15 \\\cline{1-6}
 \end{tabular}
 \vspace{0.8cm}
\caption{Parameters appearing in the fit function (i) for form
factors of the $B_{u}\rightarrow D_{u}^{\ast}(2007)\ell\nu$ at
$M_{1}^2=19~GeV^2$, $M_{2}^2=5~GeV^2.$} \label{tab:1}
\end{table}

\begin{table}[h]
\centering
\begin{tabular}{|c||c|c|c|c|c|} \hline
  & f(0)  & $\alpha$ & $\beta$& $\gamma$& $\lambda$\\\cline{1-6}\hline \hline
 $f_{V}$ & 0.47 & -3.08 & 4.83& -5.95& 2.95\\\cline{1-6}
 $f_{0}$ & 0.24 & -2.20 & 2.18& -1.83& -1.90\\\cline{1-6}
 $f_{+}$ & 0.14  & -4.13 & 8.99& -15.10& 10.65\\\cline{1-6}
 $f_{-}$ & -0.16  & -3.87 & 7.73& -12.71& 8.26\\\cline{1-6}
 \end{tabular}
 \vspace{0.8cm}
\caption{Parameters appearing in the fit function (i) for form
factors of the $B_{d}\rightarrow D_{d}^{\ast}(2010)\ell\nu$ at
$M_{1}^2=19~GeV^2$, $M_{2}^2=5~GeV^2.$} \label{tab:1}
\end{table}

\begin{table}[h]
\centering
\begin{tabular}{|c||c|c|c|} \hline
  & a  & b & $m_{fit}^{2}$\\\cline{1-4} \hline \hline
 $f_{V}$ & 55.03 & -54.30 & 23.18\\\cline{1-4}
 $f_{0}$ & 1.43  & -4.32 & 18.80\\\cline{1-4}
 $f_{+}$ & 1.14  & -2.57 & 14.88\\\cline{1-4}
 $f_{-}$ & -2.80  & 3.43 & 14.60\\\cline{1-4}
 \end{tabular}
 \vspace{0.8cm}
\caption{Parameters appearing in the fit function (ii) for form
factors of the $B_{s}\rightarrow D_{s}^{\ast}(2112)\ell\nu$ at
$M_{1}^2=19~GeV^2$, $M_{2}^2=5~GeV^2.$} \label{tab:1}
\end{table}

\begin{table}[h]
\centering
\begin{tabular}{|c||c|c|c|} \hline
  & a  & b & $m_{fit}^{2}$\\\cline{1-4} \hline \hline
 $f_{V}$ & 118.69 & -108.48 & 23.43\\\cline{1-4}
 $f_{0}$ & 4.54  & -5.12 & 20.74\\\cline{1-4}
 $f_{+}$ & 7.79  & -5.84 & 14.57\\\cline{1-4}
 $f_{-}$ & -6.72  & 5.46 & 14.02\\\cline{1-4}
 \end{tabular}
 \vspace{0.8cm}
\caption{Parameters appearing in the fit function (ii) for form
factors of the $B_{u}\rightarrow D_{u}^{\ast}(2007)\ell\nu$ at
$M_{1}^2=19~GeV^2$, $M_{2}^2=5~GeV^2.$} \label{tab:1}
\end{table}

\begin{table}[h]
\centering
\begin{tabular}{|c||c|c|c|} \hline
  & a  & b & $m_{fit}^{2}$\\\cline{1-4} \hline \hline
 $f_{V}$ & 115.74 & -106.73 & 23.41\\\cline{1-4}
 $f_{0}$ & 10.43  & -12.85 & 20.66\\\cline{1-4}
 $f_{+}$ & 5.50  & -5.07 & 14.58\\\cline{1-4}
 $f_{-}$ & -5.36  & 4.90 & 14.03\\\cline{1-4}
 \end{tabular}
 \vspace{0.8cm}
\caption{Parameters appearing in the fit function (ii) for form
factors of the $B_{d}\rightarrow D_{d}^{\ast}(2010)\ell\nu$ at
$M_{1}^2=19~GeV^2$, $M_{2}^2=5~GeV^2.$} \label{tab:1}
\end{table}

In deriving the numerical values for the ratio of the form factors
at HQET limit, we take the value of the $\Lambda$ and
$\overline{\Lambda}$ obtained from two point sum rules,
$\Lambda=0.62 GeV$ \cite{huang} and $\overline{\Lambda}=0.86
GeV$\cite{dai}. The following relations are defined for the ratio of
the form factors,
\begin{eqnarray}\label{rler}
 R_{1(2)[3]}&=&\Bigg[1-\frac{q^{2}}{(m_{B}+m_{D^{*}})^{2}}\Bigg]\frac{f_{V(+)[-]}(y)}{f_{0}(y)},
\nonumber\\
  R_{4(5)}&=&\Bigg[1-\frac{q^{2}}{(m_{B}+m_{D^{*}})^{2}}\Bigg]\frac{f_{+(-)}(y)}{f_{V}(y)},
  \nonumber\\
  R_{6}&=&\Bigg[1-\frac{q^{2}}{(m_{B}+m_{D^{*}})^{2}}\Bigg]\frac{f_{-}(y)}{f_{+}(y)},\nonumber\\
 \end{eqnarray}

The numerical values of the above mentioned ratios and a comparison
of our results with the predictions of \cite{neubert2} which
presents the application of the subleading Isgur-Wise form factors
for $B\rightarrow
 D^{\ast}\ell\nu$ are shown in Table 8. Note that the values in this
 Table are obtained with $T_{1}=T_{2}=2~ GeV$ correspond to $M_{1}^{2}=19~
 GeV^{2}$ and $M_{2}^{2}=5~ GeV^{2}$ which are used in Tables [2-7].
\begin{table}[h]
\centering
\begin{tabular}{|c||c|c|c|c|c|c|} \hline
  y& 1 (zero recoil) & $1.1$ & $1.2$& $1.3$& $1.4$&1.5\\\cline{1-7}
 $q^2 (GeV^{2})$ & 10.69 & 8.57 & 6.45& 4.33& 2.20&0.08\\\cline{1-7}
  \hline \hline
 $R_{1}$ & 1.34 & 1.31 & 1.25& 1.19& 1.10&0.95\\\cline{1-7}
 $R_{2}$ & 0.80  & 0.99 & 1.10& 1.22& 1.30&1.41\\\cline{1-7}
 $R_{3}$ & -0.80  & -0.79 & -0.80& -0.81& -0.80&-0.80\\\cline{1-7}
 $R_{4}$ & 0.50  & 0.64 & 0.77& 0.94& 1.20&1.46\\\cline{1-7}
 $R_{5}$ & -0.50  & -0.51 & -0.56& -0.62& -0.71&-0.89\\\cline{1-7}
 $R_{6}$ & -0.80  & -0.67 & -0.64& -0.61& -0.55&-0.53\\\cline{1-7}
 $R_{1}$ \cite{neubert2}& 1.31  & 1.30 & 1.29& 1.28& 1.27&1.26\\\cline{1-7}
 $R_{2}$ \cite{neubert2}& 0.90  & 0.90 & 0.91& 0.92& 0.92&0.93\\\cline{1-7}
 \end{tabular}
 \vspace{0.8cm}
\caption{The values for the $R_{i}$ and comparison of $R_{1, 2}$
values with the predictions of \cite{neubert2}.} \label{tab:4}
\end{table}

Table 8 shows a good consistency between our results and the
prediction of \cite{neubert2} for $R_{1}$ at zero recoil limit,
$y=1.1$ and $1.2$, but for the other values of y, the changes in
present work results are little greater. The values for $R_{2}$
shows an approximate agreement between two predictions, however the
changes in the value of $R_{2}$ in our work is also a bit more then
\cite{neubert2}. For both $R_{1}$ and $R_{2}$, our study  and
\cite{neubert2} predictions have the same behavior, i.e., $R_{1}$
decreases when the value of y is increased and increasing in the
value of y causes the increasing in the value of $R_{2}$. From this
Table, we also see that the $R_{4}$ is sensitive to the changes in
the value of y. However, the results of $R_{3}$, $R_{5}$ and $R_{6}$
vary slowly with respect to y. Our numerical analysis for $1/m_{b}$
corrections of form factors in Eq. (\ref{3333au}) shows that this
correction increase the HQET limit of the form factors $f_{V}$ and
$f_{+}$ about $7.1^{0}/_{0}$ and $6^{0}/_{0}$, respectively, however
it doesn't change the $f_{0}$ and decrease the $f_{-}$ about
$6.5^{0}/_{0}$.

  The next step is to calculate the differential decay width in terms of the form factors.
After some calculations for differential decay rate
\begin{eqnarray}\label{28au}
\frac{d\Gamma}{dq^2}&=&\frac{1}{8\pi^4m_{B_{q}}^2}\mid\overrightarrow{p'}\mid
G_{F}^2\mid V_{cb}\mid^2\{(2A_{1}+A_{2}q^2)[\mid
f'_{V}\mid^2(4m_{B_{q}}^2\mid\overrightarrow{p'}\mid^2)+\mid
f'_{0}\mid^2]\}\nonumber\\
&+&\frac{1}{16\pi^4m_{B_{q}}^2}|\overrightarrow{p'}|
G_{F}^2|V_{cb}|^2\left\{(2A_{1}+A_{2}q^2)\Bigg[\mid
f'_{V}\mid^2(4m_{B_{q}}^2\mid\overrightarrow{p'}\mid^2 \right.
 \nonumber \\ &+&
m_{B_{q}}^2\frac{\mid\overrightarrow{p'}\mid^2}{m_{D_{q}^{\ast}}^{2}}
(m_{B_{q}}^2-m_{D_{q}^{\ast}}^2-q^2))+\mid f'_{0}\mid^2 \nonumber \\
&-& \mid
f'_{+}\mid^2\frac{m_{B_{q}}^2\mid\overrightarrow{p'}\mid^2}{m_{D_{q}^{\ast}}^2}(2m_{B_{q}}^2
+2m_{D_{q}^{\ast}}^2 -q^2)-\mid
f'_{-}\mid^2\frac{m_{B_{q}}^2\mid\overrightarrow{p'}\mid^2}{m_{D_{q}^{\ast}}^2}q^2
\nonumber\\&-& 2 \left.
\frac{m_{B_{q}}^2\mid\overrightarrow{p'}\mid^2}{m_{D_{q}^{\ast}}^2}(Re(f'_{0}
f_{+}^{'\ast}+f'_{0}
f_{-}^{'\ast}+(m_{B_{q}}^2-m_{D_{q}^{\ast}}^2)f'_{+}f_{-}^{'\ast}))\right]
\nonumber \\ &-&
2A_{2}\frac{m_{B_{q}}^2\mid\overrightarrow{p'}\mid^2}{m_{D_{q}^{\ast}}^2}
\Bigg[\mid f'_{0}\mid^2+(m_{B_{q}}^2-m_{D_{q}^{\ast}}^2)^2\mid
f'_{+}\mid^2+q^4\mid f'_{-}\mid^2  \nonumber \\ &+& 2(
\left.m_{B_{q}}^2-m_{D_{q}^{\ast}}^2)Re(f'_{0}f_{+}^{'\ast})
+2q^2f'_{0}f_{-}^{'\ast}+2q^2(m_{B_{q}}^2-m_{D_{q}^{\ast}}^2)Re(f'_{+}f_{-}^{'\ast})
\Bigg]\right\},
\nonumber \\
\end{eqnarray}
is obtained,
where\\
\begin{eqnarray}\label{30au}
\mid\overrightarrow{p'}\mid&=&\frac{\lambda^{1/2}(m_{B_{q}}^2,m_{D_{q}^{\ast}}^2,q^2)}{2m_{B_{q}}},\nonumber\\
A_{1}&=&\frac{1}{12q^2}(q^2-m_{l}^2)^2I_{0},\nonumber\\
A_{2}&=&\frac{1}{6q^4}(q^2-m_{l}^2)(q^2+2m_{l}^2)I_{0},\nonumber\\
I_{0}&=&\frac{\pi}{2}(1-\frac{m_{l}^2}{q^2}),\nonumber\\
f_{0}'&=&f_{0} (m_{D_{q}^{*}}+m_{B_{q}}),\nonumber\\
f_{V}'&=& \frac{f_{V}}{(m_{D_{q}^{*}}+m_{B_{q}})},\nonumber\\
f_{+}'&=& \frac{f_{+}}{(m_{D_{q}^{*}}+m_{B_{q}})},\nonumber\\
f_{-}'&=& \frac{f_{-}}{(m_{D_{q}^{*}}+m_{B_{q}})}.
\end{eqnarray}

 The following part presents evaluation of the value of the branching ratio of these decays.
 Taking into account the $q^2$ dependence of
the form factors and performing integration over $q^2$ in the
interval $m_{l}^2\leq q^2\leq(m_{B_{q}}-m_{D_{q}^{\ast}})^2$ and
using the total life-times $\tau_{B_{u}}=1.638\times10^{-12}s$ ,
$\tau_{B_{d}}=1.53\times10^{-12}s$ \cite{Yao} and
$\tau_{B_{s}}=1.46\times10^{-12}s$ \cite{26}, the branching ratios
which are the same for both fit functions are obtained as:
\begin{eqnarray}\label{31au}
\textbf{\emph{B}}(B_s\rightarrow
D_{s}^{\ast}(2112)\ell\nu)&=&(1.89-6.61) \times 10^{-2},\nonumber\\
\textbf{\emph{B}}(B_d\rightarrow
D_{d}^{\ast}(2010)\ell\nu)&=&(4.36-8.94) \times 10^{-2},\nonumber\\
\textbf{\emph{B}}(B_u\rightarrow
D_{u}^{\ast}(2007)\ell\nu)&=&(4.57-9.12) \times 10^{-2}.
\end{eqnarray}

The ranges appearing in the above equations are related to the
different lepton masses $(m_{e}, m_{\mu}, m_{\tau})$ as well as
 the errors in the value of input parameters. Finally, we would like to
compare our results of the branching ratios with the predictions of
CQM model \cite{zhao} and existing experimental data in Table 9.
From this Table, we see a good agreement among the phenomenological
models and the experiment for $u$ and $d$ cases. However for s case
our results are about 1.7 times smaller than that of the CQM model.
Also, there is a same behavior between present work results and the
experiment. In the experiment, the value for branching ratios
decreases from u to d. In our results also, this value decreases
from u to s cases. The order of the branching fraction in present
work for $B_{s}\rightarrow D^{\ast}_{s}\ell\nu$ decay shows that
this transition could also be detected at LHC in the near future.
For the present and future experiments about the semileptonic
$b\rightarrow c l \nu$ based decays see
~\cite{Aubert}--\cite{Drutskoy} and references therein. The
comparison of results from the experiments and phenomenological
models like QCD sum rules could give useful information about the
strong interaction inside the $D_{s}^{*}$ and its structure.

 In
conclusion, the form factors related to the $B_{q}\rightarrow
D^{\ast}_{q}\ell\nu$
 decays were calculated using
QCD sum rules approach. The HQET limit of the form factors  as well
as $1/m_{b}$ corrections to those limits were also obtained. A
comparison of the results of form factors in HQET limit with the
application of the subleading Isgur-Wise form factors at zero recoil
limit and others values of y was presented. Taking into account the
$q^{2}$ dependencies of the form factors, the total decay width and
branching ratio for these decays were evaluated. Our results are in
good agreement with that of the CQM model and existing experimental
data. The result of $B_{s}\rightarrow D^{\ast}_{s}\ell\nu$
 case shows a possibility to
detect this decay channel at LHC in the near future.
\begin{table}[h]
\centering
\begin{tabular}{|c||c|c|c|} \hline
& $B_{s}\rightarrow D_{s}^{\ast}\ell \nu $ & $B_{d}\rightarrow
D_{d}^{\ast}\ell \nu$ & $B_{u}\rightarrow D_{u}^{\ast}\ell \nu $
\\\cline{1-4}\hline\hline
Present study & $(1.89-6.61) \times 10^{-2}$ & $(4.36-8.94) \times
10^{-2}$ & $(4.57-9.12) \times 10^{-2}$\\\cline{1-4} CQM model &
$(7.49-7.66) \times 10^{-2}$ & $(5.9-7.6) \times 10^{-2}$ &
$(5.9-7.6) \times 10^{-2}$\\\cline{1-4} Experiment & - & $(5.35 \pm
0.20) \times 10^{-2}$ & $(6.5 \pm 0.5) \times 10^{-2}$\\\cline{1-4}
\end{tabular}
\vspace{0.8cm} \caption{Comparison of the branching ratio
 of the $B_{q}\rightarrow D^{\ast}_{q}\ell\nu$} decays in
present study, the CQM model \cite{zhao} and the experiment
\cite{Yao}. \label{tab:2}
\end{table}
\section{Acknowledgment}
  The authors would like to thank T. M. Aliev and A. Ozpineci for
  their useful discussions and also TUBITAK, Turkish Scientific and Research
Council, for their financial support provided under the project
103T666.
  
\clearpage
\begin{figure}
\vspace*{-1cm}
\begin{center}
\includegraphics[width=10cm]{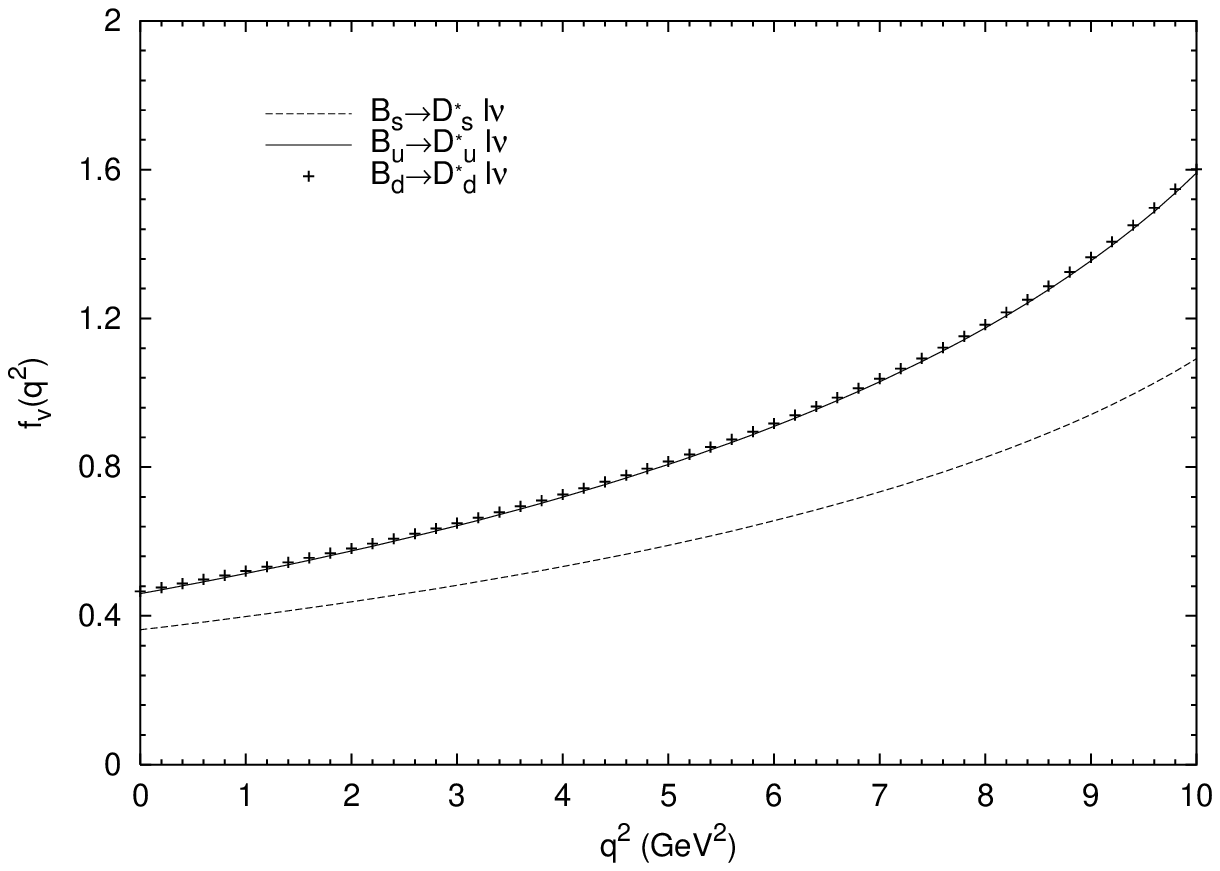}
\end{center}
\caption{The dependence of $f_{V}$ on
 $q^2$ at $M_{1}^2=19~GeV^2$, $M_{2}^2=5~GeV^2$, $s_{0}=35~GeV^2$ and $s_{0}'=6~GeV^2$. } \label{fig1}
\end{figure}
\begin{figure}
\begin{center}
\includegraphics[width=10cm]{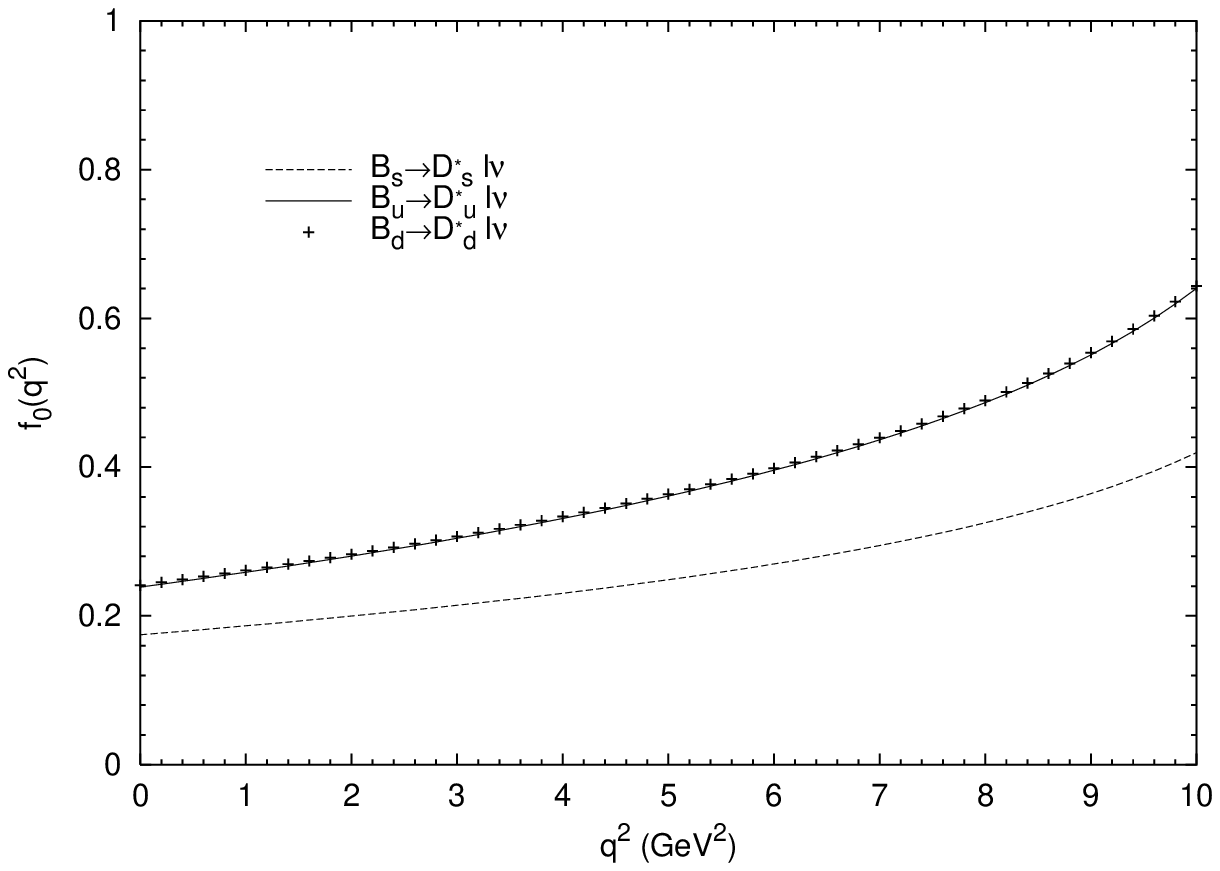}
\end{center}
\caption{ The dependence of $f_{0}$ on
 $q^2$ at $M_{1}^2=19~GeV^2$, $M_{2}^2=5~GeV^2$, $s_{0}=35~GeV^2$ and $s_{0}'=6~GeV^2$.} \label{fig2}
\end{figure}
\newpage
\begin{figure}
\vspace{-2cm}
\begin{center}
\includegraphics[width=10cm]{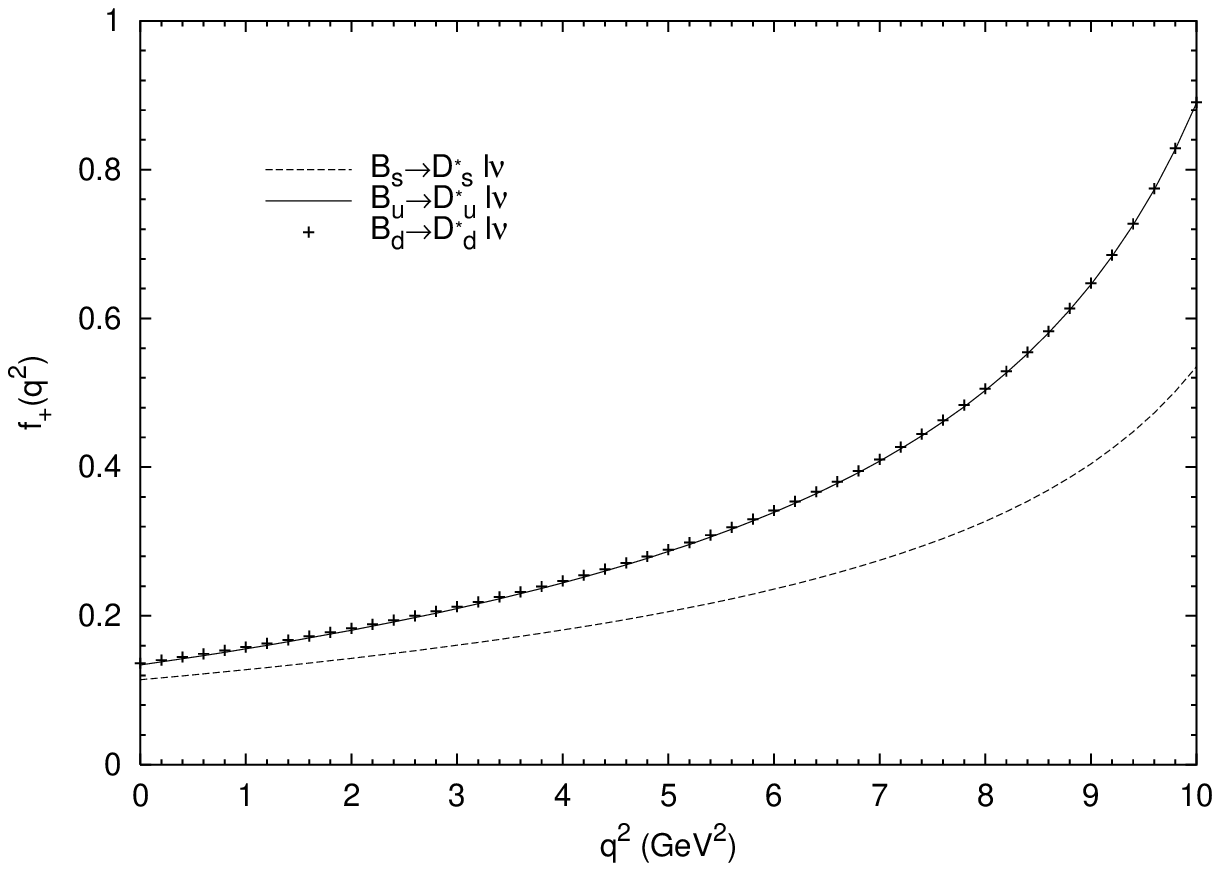}
\end{center}
\caption{The dependence of $f_{+}$ on
 $q^2$ at $M_{1}^2=19~GeV^2$, $M_{2}^2=5~GeV^2$, $s_{0}=35~GeV^2$ and $s_{0}'=6~GeV^2$.} \label{fig3}
\end{figure}
\begin{figure}
\begin{center}
\includegraphics[width=10cm]{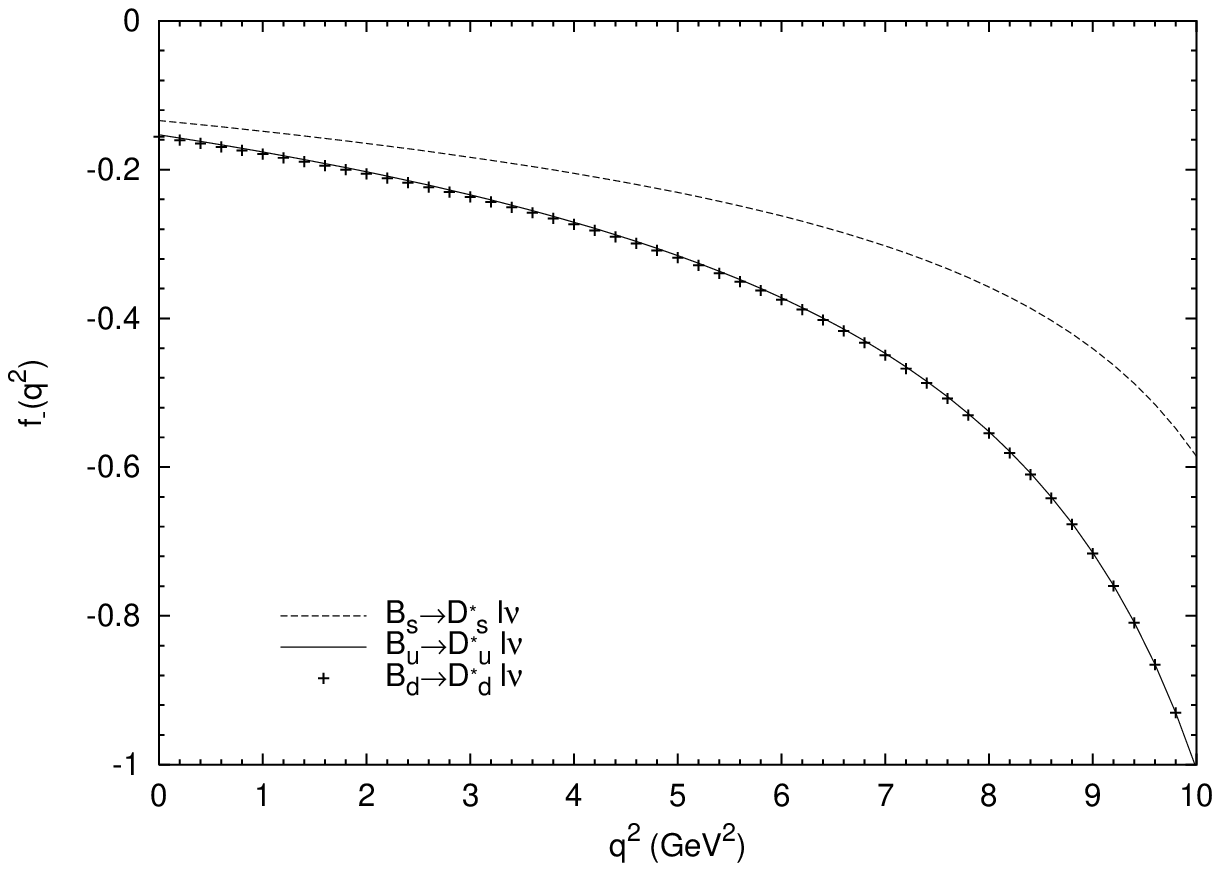}
\end{center}
\caption{The dependence of $f_{-}$ on
 $q^2$ at $M_{1}^2=19~GeV^2$, $M_{2}^2=5~GeV^2$, $s_{0}=35~GeV^2$ and $s_{0}'=6~GeV^2$.} \label{fig4}
\end{figure}
\begin{figure}
\begin{center}
\includegraphics[width=10cm]{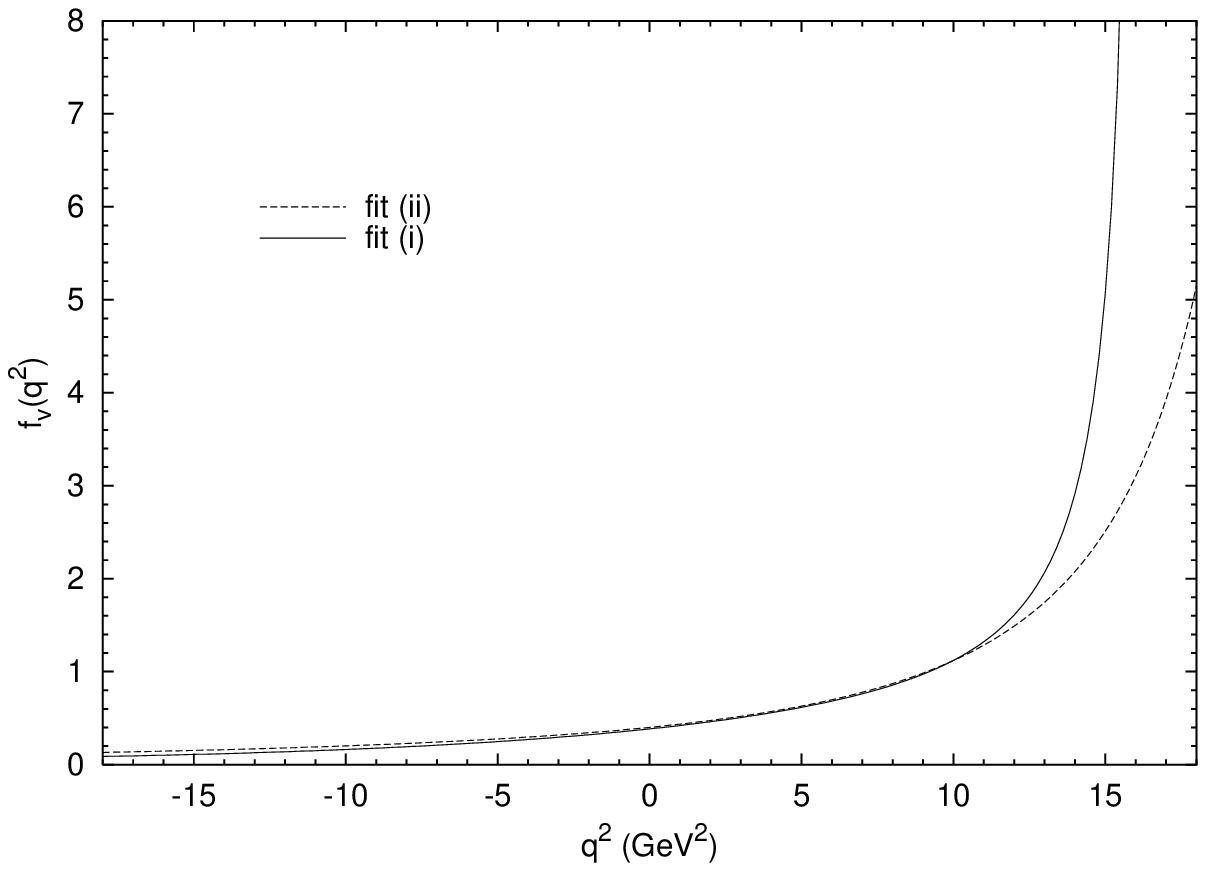}
\end{center}
\caption{Comparison of fit functions (i) and (ii) for form factor
$f_{V}$ for $q=s$. }\label{fig6}
\end{figure}

\newpage
\begin{thebibliography}{99}
\bibitem{1}  B. Aubert et. al., BaBar
Collaboration, Phys. Rev. Lett. {\bf 90} (2003) 242001.
\bibitem{2}  D. Besson et. al., CLEO Collaboration, Phys. Rev. {\bf D68} (2003) 032002.
\bibitem{3} Y. Nikami et. al., Belle Collaboration, Phys. Rev.
Lett. 92 (2004) 012002.
\bibitem{4}  P. Krokovny et. al., Belle Collaboration, Phys. Rev. Lett. {\bf 91}
  (2003) 262002 .
\bibitem{5} A. Drutskoy et. al., Belle Collaboration, Phys. Rev. Lett.
94
 (2005) 061802.
\bibitem{6} B. Aubert et. al., Babar Collaboration, Phys. Rev. Lett. 93
 (2004) 181801.
\bibitem{7} B. Aubert et. al., Babar Collaboration, Phys. Rev. D. 69
 (2004) 031101.
\bibitem{8} B. Aubert et. al., Babar Collaboration, arXiv:
0408067 [hep-ph].
\bibitem{9} P. Colangelo, F. De Fazio and R. Ferrandes, Mod. Phys.
Lett. A 19 (2004) 2083.
\bibitem{10}  E. S. Swanson, Phys. Rept. {\bf 429} (2006) 243.
\bibitem {11} P. Colangelo, F. De Fazio, and
A. Ozpineci, Phys. Rev. {\bf D72}  (2005) 074004.
 \bibitem {13}  P. Colangelo and A.
Khodjamirian, in At the Frontier of Particle Physics/Handbook of
QCD, edited by M. Shifman (World Scientific, Singapore, 2001), Vol.
3, p. 1495.
\bibitem {zhao} Shu-Min Zhao, Xiang Liu, Shuang-Jiu Li, Eur. Phy. J. {\bf C51} (2007) 601.
\bibitem{Yao} W.M. Yao et al., Particle Data Group, J. Phys. {\bf G33} (2006) 1.
\bibitem {neubert2}  M. Neubert, Phys. Rev. {\bf D46} (1992) 3914.
\bibitem {grozin1}  V. N. Baier, A. G. Grozin, Z. Phys. {\bf C47} (1990) 669.
\bibitem {ovcinkov} A. A. Ovchinnikov, V. A. Slobodenyuk, Z. Phys. {\bf C44} (1989) 433.
\bibitem{15}   P. Ball,
V. M. Braun, H. G. Dosch, Phys. Rev. {\bf D44} (1991) 3567.
\bibitem {ming} Ming Qiu Huang, Phys. Rev. {\bf D69} (2004) 114015.
\bibitem {neubert1} M. Neubert, Phys. Rep. {\bf 245} (1994) 259.
\bibitem {kazem} T. M. Aliev, K. Azizi, A. Ozpineci, Eur. Phys. {\bf
C51} (2007) 593.
\bibitem {grozin}  V. N. Baier, A. G. Grozin, In Zvenigorod 1993, Proceedings,
High energy physics and quantum field theory, Physics at VLEPP
35-41, arxiv:hep-ph/9908365.
\bibitem{20} B. L. Ioffe, Prog. Part. Nucl. Phys.
{\bf 56} (2006) 232.
\bibitem{12} M. A. Shifman, A. I. Vainshtein, and V. I. Zakharov, Nucl. Phys.
{\bf B147}  (1979) 385.
\bibitem{16}  P. Ball, Phys. Rev. {\bf D48}  (1993) 3190.
\bibitem{damir}  D. Becirevic, A. B. Kaidalov , Phys. Lett. {\bf B478}  (2000) 417.
\bibitem{damirinbali}  P. Ball, R. Zwicky, Phys. Rev. {\bf D71}  (2005) 014015.
\bibitem{damirinbraunu} V. M. Belyaev, V. M. Braun, A. Khodjamirian and R.
Ruckl, Phys. Rev. {\bf D51}  (1995) 6177.
\bibitem{huang} T. Huang, C. W. Luo, Phys. Rev. {\bf D50} (1994) 5775.
\bibitem{dai} Y. B. Dai, C. S. Huang, C. Liu, and S. L. Zhu, Phys.
Rev. {\bf D68}, 114011 (2003).
\bibitem {26} S. Eidelman et. al., Particle
Data Group, Phys. Lett. {\bf B592} (2004) 1.
 \bibitem {Aubert} B. Aubert, et. al., BABAR Collaboration,
 Phys. Rev. Lett. {\bf 100} (2008) 021801.
 \bibitem {chen} K. F. Chen,  Belle Collaboration, In the Proceedings of 5th Flavor Physics
 and CP Violation Conference (FPCP 2007), arXiv:hep-ex/0708.4089.
 \bibitem {Aubert2} B. Aubert, et. al., BABAR Collaboration, Phys. Rev. {\bf D74} (2006) 092004.
  \bibitem {Aubert3} B. Aubert, et. al., BABAR Collaboration, BABAR-CONF-04-31, 32nd
  International Conference on High-Energy Physics (ICHEP 04),
  arXiv:hep-ex/0409047.
  \bibitem {belle1}  A. Matyja, M. Rozanska, et. al., Belle
  Collaboration, Phys. Rev. Lett. {\bf 99} (2007) 191807.
 \bibitem {Drutskoy} A. Drutskoy, The proceedings of International Europhysics Conference on
 High Energy Physics (EPS-HEP 2007), arXiv:hep-ex/0710.1647.

\end{thebibliography}
\end{document}